\documentclass[aps,prl,twocolumn,showpacs]{revtex4}

\usepackage{graphicx}
\DeclareGraphicsExtensions{.png,.jpg,.eps}

\usepackage{xcolor}
\usepackage{amsmath}

\begin{document}

\title{Magnetic edge  anisotropy   in graphene-like  honeycomb crystals}

\author{J. L. Lado, J. Fern\'andez-Rossier\footnote{On leave from Departamento de F\'isica Aplicada, Universidad de Alicante,  Spain} }
\affiliation{  International Iberian Nanotechnology Laboratory (INL),
Av. Mestre Jos\'e Veiga, 4715-330 Braga, Portugal
}

\author{}

\date{\today} 

\begin{abstract}

The  independent predictions of edge ferromagnetism and the Quantum Spin Hall phase in graphene have inspired the quest of 
other two dimensional honeycomb systems, such as silicene,  germanene, stanene, iridiates, 
and  organometallic lattices, as well as artificial  superlattices,  all of them  with electronic properties analogous
to those of graphene,  but much larger spin-orbit coupling.
Here we study  the  interplay of ferromagnetic order  and spin-orbit interactions at the zigzag edges of these graphene-like systems.  We find   an  in-plane magnetic anisotropy that opens a gap in the otherwise conducting edge channels, that should result  in  large changes of electronic properties upon rotation of the magnetization.

\end{abstract}
\maketitle

Magnetic anisotropy, a technologically crucial property, is driven by  spin-orbit interaction, which is normally the underdog in the competition with the other two terms that control ferromagnetism, namely,  kinetic and Coulomb energy\cite{Coey:1261177}.
   As a result,  magnetic anisotropy energy  in conventional ferromagnets is at least 2 orders of magnitude smaller than  the Curie temperature and the Fermi energies (or the  band-gap, in the case of insulators).  For the same reason, transport properties in ferromagnetic metals  are only weakly dependent on the magnetic orientation, and typical values for the anisotropic magnetoresistance (AMR) are below 3 percent\cite{Coey:1261177}. 

Here we study magnetic anisotropy in a class of systems for which the balance between these three energy scales is very different from the usual, which leads to two dramatic consequences, very different from conventional ferromagnetism. First,  the conducting properties  change from metal to insulator, depending on the magnetization orientation, an effect  that, to the best of our knowledge, has never been reported. Second,  the magnetic moment magnitude depends strongly on the magnetic orientation, and it can change even vanish in some directions, a phenomenon dubbed colossal magnetic anisotropy\cite{SmogunovA.2008,Nunez2012403}.    The class of systems in question are 
   the zigzag edges of two dimensional honeycomb crystals\cite{PhysRevB.54.17954}   whose electronic properties can be described with a tight-binding model with
a single orbital per site and Kane-Mele spin-orbit interactions\cite{kane-mele}. 
    This includes several materials, such as  group IV two dimensional crystals
(graphene\cite{kane-mele}, as silicene\cite{silicene1,silicene3,gmrsilicene}, 
germanene\cite{Houssa2010}, and stanene\cite{honey-tin}), 
the double layer perovskyte iridates \cite{PhysRevLett.102.256403,perovs-t2g5},   and 
 metal organic frameworks (MOF) \cite{Wang2013}. In addition, given that the existence of non-dispersive edge states occurs at the zigzag edge of  any system described with the Dirac equation\cite{PhysRevB.73.235411}, the results discussed here should also be valid for the so-called designer Dirac fermions formed  in "artificial graphene" formed by decoration of two dimensional electron gases with honeycomb arrangements\cite{Gomes2012,art-honey}.

Ignoring spin-orbit and Coulomb interactions altogether, these 2D crystals are zero band-gap semiconductors with Dirac-like dispersion close to the Fermi energy.  Zigzag edges in these systems are known to host localized edge states that, when both Coulomb and spin-orbit coupling are neglected, are non-dispersive, sub lattice polarized,  and lie precisely at the Fermi energy, at half-filling\cite{PhysRevB.54.17954}. The ensuing large density of states results in a Stoner  instability that leads to  ferromagnetic order at the edge\cite{FujitaJPSJ96,Son06,JFR07,JFR08}.    

On the other hand,  Kane-Mele spin-orbit interaction, a second-neighbor spin dependent hopping that conserves the spin component $s_z$ perpendicular to the two dimensional crystal\cite{kane-mele}, has dramatic consequences in these honeycomb crystals.  It opens a topologically non-trivial gap in bulk and  the emergence of  in-gap spin-filtered dispersive edge states: 
 for a given spin projection $s_z$, electrons propagate along one direction only, preventing back-scattering even in the presence of time-reversal symmetric disorder.  Importantly, the slope of the edge bands is proportional to the Kane-Mele spin-orbit coupling, which controls thereby  the density of states at the Fermi energy.    The interplay of spin-orbit and Coulomb repulsion on the otherwise non-dispersive edge states leads to the strong magnetic anisotropy effects anticipated above.

\begin{figure}
 \centering
                \includegraphics[width=0.5\textwidth]{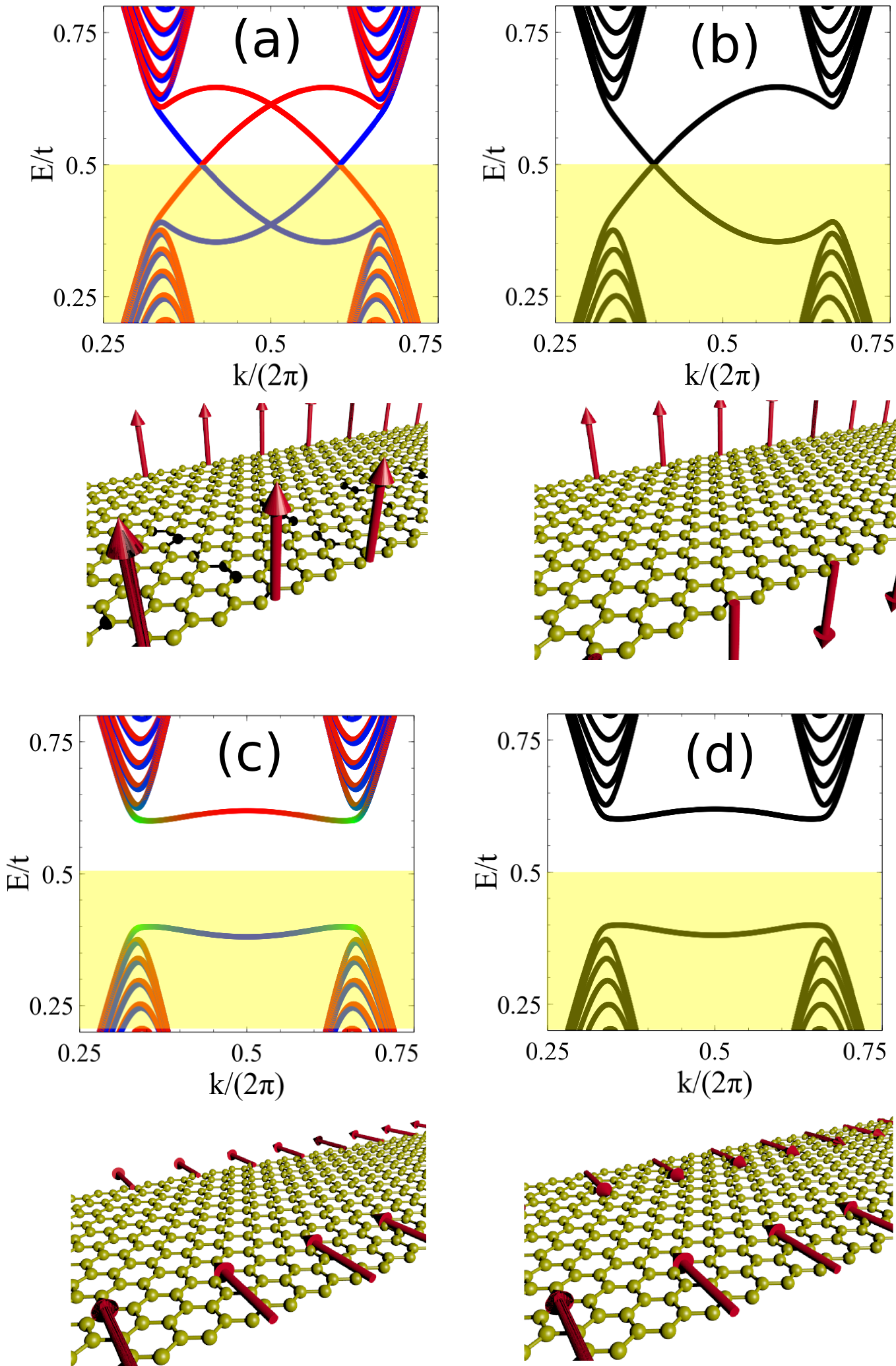}
\caption{Four different ferromagnetic configurations, either in and off-plane and parallel (FM) and antiparallel (AF) edge magnetization, together with their band structures calculated within the mean field Kane-Mele-Hubbard model.  
(a,b) Off plane and conducting (both for FM and AF arrangements).  (c,d) In-plane and insulating parallel (both FM and AF) magnetizations.
Calculations done with  $U=t$ and  $t_{SO}=0.02t$ }
\label{f1}
\end{figure}

To model this kind of systems, we use the so
called Kane-Mele-Hubbard Hamiltonian\cite{Rachel-2010,Soriano-JFR10}, which
provides a minimal model to study the effect of the Coulomb interactions on the topologically protected edge states: 
\begin{equation}
H=\sum_{\langle ij \rangle,\sigma} t c^\dagger_{i\sigma} c_{j\sigma}+\sum_{\langle\langle ij \rangle \rangle,\sigma } i
 t_{SO} \sigma \nu_{ij} c^\dagger_{i\sigma} c_{j\sigma}  + H_{\rm int}
\end{equation}
where $\sigma=\pm 1$ are the spin projections of the spin along the axis perpendicular to the 
to the two dimensional crystal, $\langle\rangle$ stands for first neighbor and $\langle\langle\rangle\rangle$ for second,
  $\nu_{ij}=\pm 1$ for clockwise or anti-clockwise second
neighbor hopping \cite{haldane,kane-mele}.

For simplicity, we neglect the Rashba coupling,\cite{kmU-rashba,rashba-si}. In the case of planar honeycomb systems, such as graphene, the Rashba term is null. For buckled group IV crystals, such as silicene, germanene and stanene,   the magnitude of the Rashba is one order of magnitude lower than the
pure spin-orbit.\cite{group-iv}

The Hubbard term reads:
\begin{equation}
H_{int}= U \sum_i n_{i\uparrow}n_{i\downarrow}
\end{equation}
where $n_{i\uparrow}=c^{\dagger}_{i\uparrow}c_{i\uparrow}$ denotes the occupation operator of site $i$ with spin $\uparrow$ along an arbitrary quantization axis.  We treat the Hubbard interaction in the collinear mean field approximation, enforcing the magnetization to lie along the axis  $\vec\Omega=(sin\alpha,0,cos\alpha)$, that we take as the quantization axis (see Fig.\ref{f2}a).  This approach permits to study solutions with different $\alpha$ and compare their properties.   Rotations in the $xy$ plane leave the results invariant, due to the symmetry of the Kane-Mele spin-orbit coupling.
In general, the Coulomb interaction term evaluated in the mean field approximation leads to two self-consistent potentials terms, direct and exchange. In the case of the Hubbard model in the collinear approximation, only the direct term survives:
\begin{equation}
H_{MF}=U[\langle n_{i\uparrow (\alpha)}\rangle n_{i\downarrow (\alpha)}+
n_{i\uparrow (\alpha)}\langle n_{i\downarrow (\alpha)} \rangle-
\langle n_{i\uparrow (\alpha)}\rangle\langle n_{i\downarrow (\alpha)}\rangle]
\end{equation}
where the notation explicitly shows the spin quantization axis is taken along $\vec{\Omega}(\alpha)$ and 
$\langle n_{i\uparrow (\alpha)}\rangle$ stand for the average of the occupation operator calculated within the ground state of the mean field Hamiltonian:
\begin{equation}
H=H_0+H_{MF}
\end{equation}
As usual, this defines a self-consistent problem that we solve by iteration.  
Because of the  spin-orbit Kane-Mele term in $H_0$,    mean field solutions with different $\alpha$ are  not equivalent.  Notice as well that the $H_{MF}$ term is non-diagonal when represented in the basis of eigenstates of $S_z$ and $\alpha\neq 0$\cite{supplemental}. 

We pay special attention to the atomic  magnetization, along the $\vec{\Omega}(\alpha)$  in site $i$:
\begin{equation}
m_{i(\alpha)}=g \mu_B\frac{[\langle n_{i\uparrow (\alpha)}\rangle
-\langle n_{i\downarrow (\alpha)}\rangle]}{2}
\end{equation}
and we take $g=2$. 

In order to study the zigzag edges it is convenient to study  ribbons, that define a one dimensional crystal (see Fig. 1) with two edges.  A given unit cell of the one dimensional crystal is formed by $N$ units of 4 atoms.  In  the following we characterize the  width of the ribbons by $N$.     For finite $U$, and as  long as $t_{SO}/U$ is not too large,  we find solutions with ferromagnetic order at the edges. The magnetic moment calculated self-consistently is non-negligible only at the edge atoms. 
Attending to their mutual magnetization orientation, ribbons yield two types of solutions with ferromagnetic edges:  parallel (FM) and antiparallel (AF).   For sufficiently wide ribbons the inter-edge coupling is negligible and both solutions have identical properties. 

 The first important   result of the paper is shown in figure 1.   Whereas off-plane magnetization $(\alpha=0)$ leads to a conducting solution, found in previous works\cite{Soriano-JFR10},  the in-plane magnetization opens a gap.   Therefore, transport properties of zigzag edges will change dramatically upon rotation of the magnetization direction, in contrast with conventional metallic ferromagnets.
This metal-insulator transistion
will is developed as well in chiral edge ribbons \cite{supplemental}, which have
been widely reported \cite{yazyev-chiral,yazyev-acs}.

The second important result of the manuscript is shown in Fig.\ref{f2}b. The ground state energy $E(\alpha)$ is minimal for $\alpha=\frac{\pi}{2}$, i.e., for in-plane magnetization, which means that spontaneous magnetic order in this system leads to insulating behavior.

%


\begin{figure}
 \centering
                \includegraphics[width=0.5\textwidth]{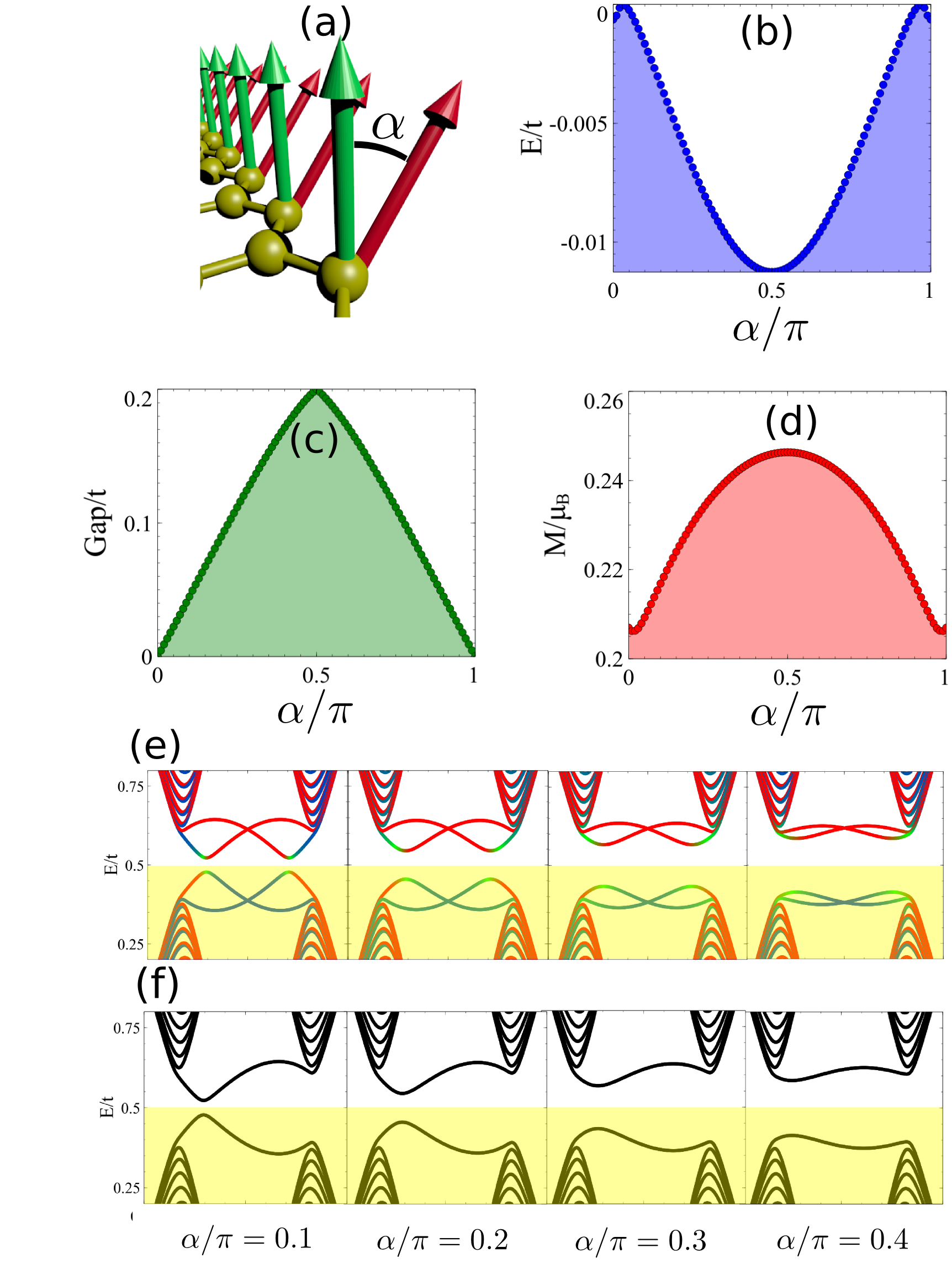}
\caption{Evolution of electronic properties for the FM ribbon as a function of the
magnetization direction $\vec{\Omega}=(sin\alpha,0,cos\alpha)$. 
(a) Scheme of the edge magnetization for two different angles $0$ and  $\alpha$.   Ground state total energy (per unit cell, with 2 magnetic atoms per cell) (b),
gap(c) and magnetization (d)  as a function of $\alpha$.
 (e) and (f) Evolution
of the band structure for different values of $\alpha$ for the FM  (e) and AF(f)  configurations.
 $N=30$, U=t, $t_{SO}=0.02t$.}
\label{f2}
\end{figure}


The results of figure 1 can be understood as follows.  In the absence of magnetic order, two spin-filtered in-gap edge states with opposite velocities exist at each edge\cite{kane-mele}, resulting in a two-fold degeneracy (not shown).   Ferromagnetic order with $\vec{\Omega}=\hat{z}$ breaks time reversal symmetry but  does not mix spins. Thus, magnetic order merely yields a spin-dependent shift  that breaks the two-fold degeneracy, as seen in Fig.\ref{f1}a, for the FM case.  For the AF configuration, there is an extra symmetry that restores the double degeneracy:  the combined action of  spatial inversion, that  results in a exchange of the atoms of the two interpenetrating triangular sub lattices $A$ and $B$ that form the honeycomb,  and time reversal, that exchange $\uparrow$ and $\downarrow$,  leave the system invariant.  Thus the spin $\uparrow$ band localized at the at the $A$ type edge is degenerate with the spin $\downarrow$ band localized in the opposite edge. 

The situation is radically different when the magnetization lies in plane.  Representing the self-content potential in the
basis of the $U=0$ 
spin filtered edge states, with   spin quantized  along the $\hat{z}$ axis, the effect of the in-plane magnetization is to mix bands with opposite spins. As a result,  a band gap opens at the $k$ point where the non-interacting edge bands cross.  The evolution of the bands as the magnetization is rotated from almost off-plane (left) to almost in-plane (right) is shown if Fig. \ref{f2})(e,f).  It is apparent that the band gap (Fig. \ref{f2})(c)) is maximal for in-plane magnetization ($\alpha=\frac{\pi}{2}$) and null for off-plane $\alpha=0$.  
The preference for  in-plane magnetization  can also be connected with the variation of the magnitude of the edge magnetic moment with $\alpha$ (Fig.\ref{f2}d).   These two results naturally explain the fact that the ground state energy is minimal for in-plane magnetization. At half-filling, all the valence bands are occupied and the conduction bands are empty. Therefore, increasing the band-gap decreases the total energy. 

The gap opening as long as magnetization is not off-plane will certainly have dramatic consequences on the transport properties along the edges.  A result similar to this has been obtained recently\cite{rachelezawa2013}, using a Kane-Mele model where magnetic order is externally driven, and modeled by  a magnetic exchange potential that arises from proximity rather than spontaneously, as discussed here.

\begin{figure}
 \centering
                \includegraphics[width=0.5\textwidth]{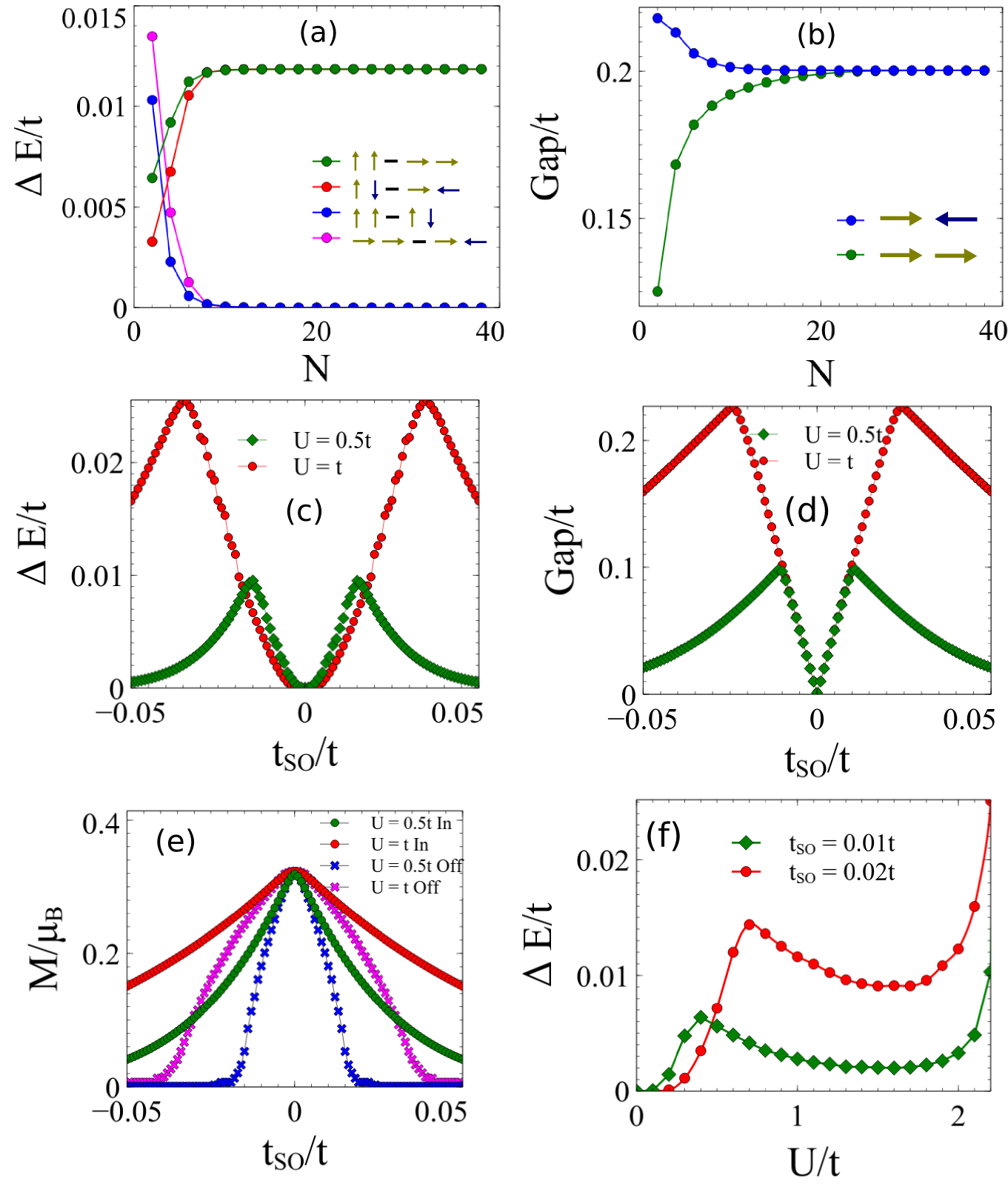}
\caption{(a) Evolution of the energy differences between the in-plane vs off-plane as well as the FM vs AF 
 configurations (4 cases) as a function of ribbon width $N$. (b) Gap, for in-plane magnetization solution, for FM and AF solutions, as a function of $N$. 
(c,d,e) Evolution with the strength of the spin-orbit coupling $t_{SO}$ : anisotropy energy (c), gap (d) and edge
magnetization (e). (f) Evolution of the anisotropy energy with the on-site Hubbard interaction. In (a-b) $U=t$ and $t_{SO}=0.02 t$.  In (c-d) $N=30$.
}
\label{f3}
\end{figure}

The results of figures 1 and 2 are for a specific choice of $U/t=1$ and $t_{SO}/t=0.02$, and for a ribbon with $N=30$ sites, wide enough to decouple the two edges.  We now discuss how the results depend, quantitatively, on the specific values of the spin-orbit coupling,   ribbon width and $U$.
The evolution of several  energy differences
between AF/FM and in-plane off-plane configurations, as a function of the ribbon width $N$ is shown in  Fig.\ref{f3}a .
For  large $N$ it is apparent that 
FM and AF have the same ground state and anisotropy energy.
In addition, the edge gap (\ref{f3}b) also becomes independent on the magnetic configuration
at large width.

\begin{table}
\centering
\begin{tabular}{lccc}
\hline
\hline
{\bf Material} &  $t_{SO}  $  & $t$ &   {\bf Reference} \\
\hline
Graphene & 0.1-5.0 $\mu$eV               &  2.7 eV  &  \cite{min2006,graphene-so,fabian-so} \\
Silicene & 0.16 meV                & 1.5 eV   &  \cite{group-iv,si-ge-so} \\
Germanene & 2.5 meV       & 1.4 eV    &  \cite{group-iv,si-ge-so} \\
Stanene & 8-30  meV                     & 1.3 eV   &  \cite{group-iv,honey-tin} \\
MOF & $1 $  meV          &  0.3 eV  &  \cite{Wang2013} \\
Double perovskyte & $1 $  meV          & 0.1 eV    &  \cite{perovs-t2g5} \\
 \hline
 \end{tabular}
 \label{table}
 \caption{Energy scales for different graphene-like honeycomb materials. }
\end{table}

In figure ( Fig.\ref{f3}(c)(e))) we plot  the dependence of the magnetic anisotropy and magnetic moment (both in and off-plane) on the magnitude of the spin-orbit coupling $t_{SO}/t$, for two different values of $U$ .
 Attending to the difference between the magnitude of the magnetic moment in the off-plane and in-plane cases (Fig.\ref{f3}(e)), three different regions are found. For  very small $t_{SO}/t$   the magnetic moment is the same for out of plane and in-plane magnetization and 
 the magnetic anisotropy energy depends quadratically on $t_{SO}/t$ .   From this standpoint, the behavior of the zigzag edge is similar to conventional magnets, although a small gap,   open for in-plane magnetization.   
For wide enough ribbons, the value of this gap is given by  $min\left(\Delta_{\rm 2D}, U m_{\rm edge}\right)$ where $\Delta_{\rm 2D}=6\sqrt{3}t_{SO}$ is the bulk gap opened by spin-orbit interaction and $U m_{\rm edge}$ is the exchange splitting gap, which is a decreasing function of $t_{SO}$, giving rise to the curve seen in  figure \ref{f3}(d).

For intermediate values of $t_{SO}/t$ it is apparent that the magnetic moment magnitude is different for in-plane and off-plane orientations, but in both cases finite.  In this region the anisotropy energy scales approximately linear with $t_{SO}/t$, and the band-gap of in-plane magnetization is still a linear function of $t_{SO}/t$.   Finally, above a given critical value $t_{SO}\simeq 0.02t$  ($0.04t$) for $U=0.5t$  ($U=t$),  the system enters in the so-called colossal\cite{SmogunovA.2008,Nunez2012403} magnetic anisotropy regime,   for which magnetic order is only possible in-plane, and the  magnetic solutions off-plane do not exist. 
 Increasing spin-orbit coupling beyond this point starts to reduce the band-gap and the magnetic order altogether, which leads to a reduction of the magnetic anisotropy energy (Fig.\ref{f3}(e) )

We thereby expect that graphene, silicene and germanene are in the small $t_{SO}/t$ region, MOF is in the intermediate region and the stanene zigzag edge could show  the colossal  magnetic anisotropy  effect.  Notice that in the intermediate region  the magnetic anisotropy energy per magnetic atom can be extremely large. For instance, for stanene, taking $U=t$, 
 intermediate $t_{SO}\simeq 8$meV and $t\simeq 1.3eV$, we obtain  of $\Delta E\simeq 4 $meV,  significantly larger than record materials such as $YCo_5$\cite{ Coey:1261177}.
 Na$_2$IrO$_3$ and related systems\cite{PhysRevLett.102.256403,perovs-t2g5}
offer also a fascinating  possibility of real tuning of the effective $t_{SO}$ by
strain,\cite{perovs-t2g5}, which would make it possible to build  devices with strain-tunable anisotropy.

Given the spread of estimates of the actual values of $U$ for a given material, as well as the fact that it different substrates can result in different values of $U$,  we address the question of how the results above depend on the strength of the on-site Coulomb repulsion $U$.   At finite $t_{SO}$ there is a critical $U_{c1}$ below which the edges are non-magnetic\cite{Soriano-JFR10}, and a second $U_{c2}$ above which the entire honeycomb lattice becomes antiferromagnetic\cite{FujitaJPSJ96}.
   For $U_{c1}<U<U_{c2}$  only  the edge is magnetic,  and its  magnetic anisotropy energy is a non-monotonic function of $U$. It increases first, reflecting the increase of the magnetic moment, and then it decreases slightly, reflecting the reduction of the ratio $t_{SO}/U$. As $U$ approaches $U\simeq 2.2 t$ the magnetic anisotropy overshoots because the bulk becomes magnetic as well.


We now discuss the physical effects not covered within our two main approximations, namely,  treating the interactions at the mean field level and ignoring the Rashba spin-orbit term.  In one dimension, collective spin fluctuations are expected to destroy the infinitely long-range order described by  mean field theory.  
 Still, for ribbons shorter than the spin correlation length $\xi(T)$  the mean field theory provides a fair description, very much like density functional theory describes properly the magnetization of clusters and nanomagnets.
The spin correlation length $\xi(T)$ in graphene edges, calculated within the spin wave approximation and ignoring spin-orbit coupling\cite{Yazyev08}, is  $\xi\simeq 40\AA$ for $T=75K$.    The magnetic anisotropy barrier to rotate the spins out of plane for the approximately 15 tin atoms of a zigzag stanene edge that long would
 be $\Delta\simeq $60 meV.  
 
 Inclusion of the Rashba coupling would have two consequences. First, lack of inversion symmetry would split the bands in the case of AF configurations. Second,  it would  break the in-plane $xy$ magnetic symmetry at the edges. 

In conclusion,  we have studied the magnetic anisotropy of the ferromagnetic phase of the zigzag edges of graphene and graphene-like systems, that can be described with a single orbital Hubbard model  model on a honeycomb lattice with spin-orbit coupling described with the Kane-Mele Hamiltonian.  This includes a large class of two dimensional crystals, such as silicene\cite{silicene1,silicene3,gmrsilicene}, germanene\cite{Houssa2010}, stanene\cite{honey-tin}, iridates\cite{PhysRevLett.102.256403,perovs-t2g5} and metal organic frameworks\cite{Wang2013}.    Since the electronic dispersion of the non-interacting edge states is fully determined by the spin-orbit coupling, the resulting magnetic anisotropy effects, computed within a mean field approximation,  turn out to be very strong: the system undergoes a metal to insulator transition when the magnetization is rotated out of the normal and for large values of $t_{SO}$ the magnetic solutions are only stable for in-plane magnetization. For all values
of the spin-orbit interaction we find that the ground state energy occurs for in-plane magnetization and the edge states are gapped.

 JFR acknowledges  financial supported by MEC-Spain (FIS2010-21883-C02-01) 
  and Generalitat Valenciana (ACOMP/2010/070), Prometeo. This work has been financially supported in part by FEDER funds.  We acknowledge financial support by Marie-Curie-ITN 607904-SPINOGRAPH.

\appendix

\pagebreak
\widetext
\begin{center}
\textbf{\large Supplemental Material}
\end{center}
\setcounter{equation}{0}
\setcounter{figure}{0}
\setcounter{table}{0}
\setcounter{page}{1}

In this supplemental material we present the mean field operators
needed to perform
an angle dependent collineal mean field Hubbard calculation, as well as
the angle dependent ground states of chiral Hubbard 
Kane-Mele honeycomb ribbons.

\begin{figure}
 \centering
                \includegraphics[width=0.5\textwidth]{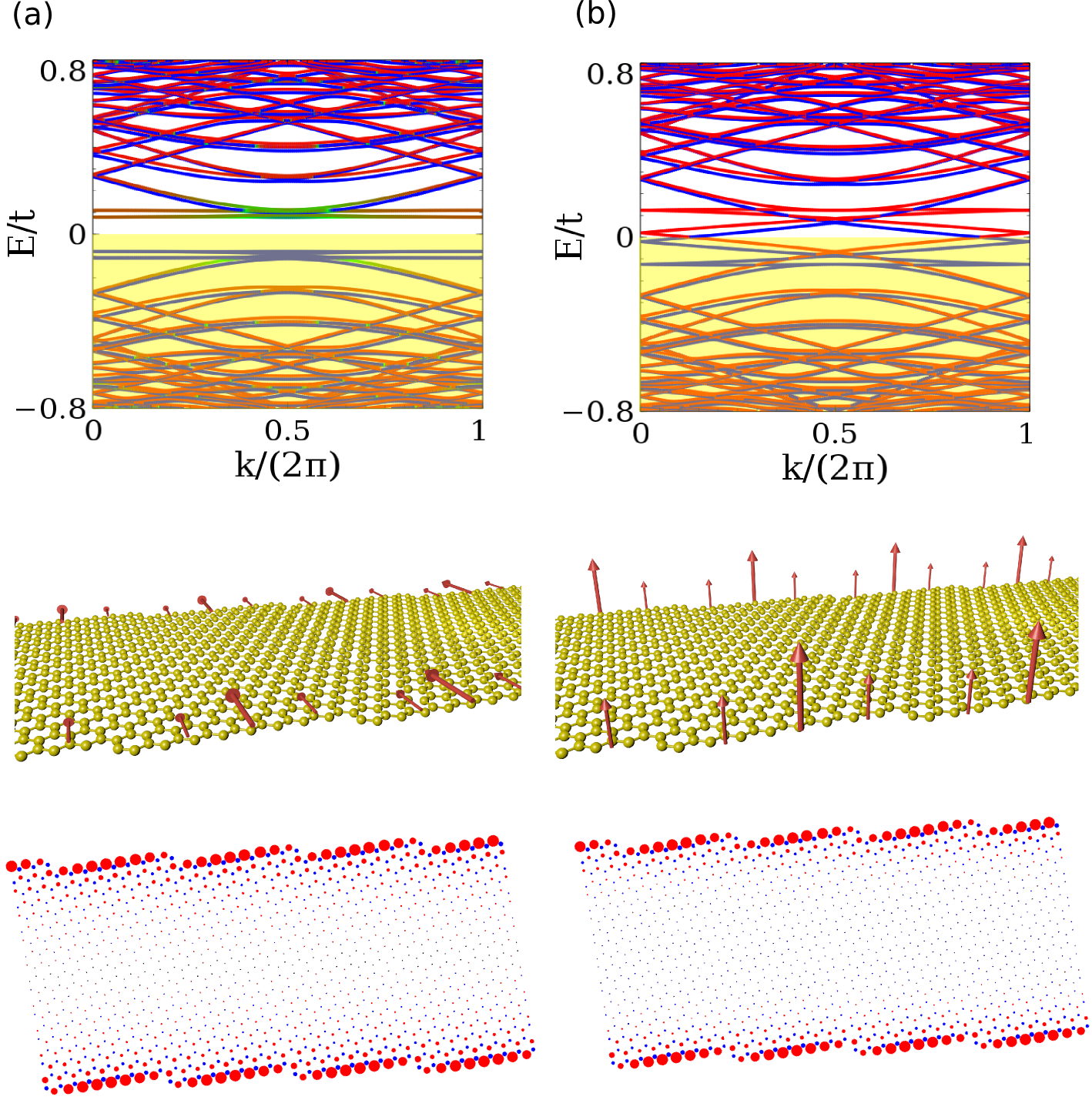}
\caption{Band structures, scheme and calculated spatial distribution
of the magnetic state in a chiral
(8,1) ribbon in edge ferromagnetic configuration for in-plane (a) and 
off-plane (b) solutions. In the same fashion as in the perfect zigzag
edge case, the in-plane magnetism opens up a gap whereas the off-plane one
remains gapless.}
\label{f2}
\end{figure}

\section{Density operators for arbitrary quantization axis}
Upon a rotation of angle $\alpha$ of the spin quantization axis
in the $xz$ plane,
the rotated of up and down density operators take the following form in the
basis of the z-axis eigenstates. 

\begin{equation}
n_{\uparrow (\alpha)} = 
\begin{pmatrix}
\cos^2\frac{\alpha}{2}  & \cos\frac{\alpha}{2}\sin\frac{\alpha}{2}\\
\cos\frac{\alpha}{2}\sin\frac{\alpha}{2} & \sin^2\frac{\alpha}{2}\\
\end{pmatrix}
\end{equation}

\begin{equation}
n_{\downarrow (\alpha)} = 
\begin{pmatrix}
\sin^2\frac{\alpha}{2}  & -\cos\frac{\alpha}{2}\sin\frac{\alpha}{2}\\
-\cos\frac{\alpha}{2}\sin\frac{\alpha}{2} & \cos^2 \frac{\alpha}{2}\\
\end{pmatrix}
\end{equation}

Thus, these are the local mean field operators in a collineal
calculation only allowing the development of magnetism along this
particular quantization axis. This choice of mean field operators
restrics the avaible Hilbert space to the collinear solutions,
whereas the inclusion of the exchange term will lead to the full
non-collinear solutions.

\section{Anisotropy in chiral ribbons}

Even though the anisosotrpy calculations were performed on a zigzag edge,
localized edge states also appear on chiral nanoribbons
\cite{yazyev-chiral,yazyev-acs}.
Thus, this anisotropy effect will appear as well in chiral
(n,m) nanoribbons
due to the local lattice imbalance. As a particular example, we show 
the selfconsistent band structure for in-plane and off-plane edge ferromagnetic
configurations for a (8,1) chiral ribbon, as reported in \cite{yazyev-chiral,yazyev-acs}.
It is observed that, in the same
fashion as in the zigzag ribbon, the off-plane magnetic solution host gapless edge states whereas the in-plane solution opens up a gap in the
topological edge states, being
this last one the lowest energy configuration of the system.

For the limiting case of pure armchair edges, magnetism is not developed
and the edge states remain gapless\cite{armchair}



\end{document}